\documentclass{aa}
\usepackage{color}
\usepackage{ulem}
\usepackage{graphicx}
\usepackage{txfonts}
\begin{document}
   \title{Stellar evolution through the ages: period variations in galactic
   RRab stars as derived from the GEOS database and TAROT telescopes}
   \author{J.F.~Le Borgne \inst{1,2} \and
           A.~Paschke \inst{1,3}  \and
           J.~Vandenbroere \inst{1} \and
           E.~Poretti \inst{1,4}  \and
           A.~Klotz \inst{5} \and
           M.~Bo\"er \inst{6} \and
           Y.~Damerdji \inst{6,7} \and
           M.~Martignoni \inst{1,3} \and
           F.~Acerbi \inst{1}
          }
   \offprints{J.F. Le Borgne}
\institute{
GEOS (Groupe Europ\'een d'Observations Stellaires), 23 Parc de Levesville, 28300 Bailleau l'Ev\^eque, France
\and
Laboratoire d'Astrophysique de Toulouse-Tarbes, Observatoire Midi-Pyr\'en\'ees (CNRS/UPS), Toulouse, France
\and
Bundesdeutsche Arbeitsgemeinschaft f\"ur Ver\"anderliche Sterne e.V. (BAV), Munsterdamm 90, D-12169 Berlin, Germany
\and
INAF-Osservatorio Astronomico di Brera, via E. Bianchi 46, I-23807 Merate, Italy
\and
Centre d'Etude Spatiale des Rayonnements, Observatoire Midi-Pyr\'en\'ees (CNRS/UPS), Toulouse, France
\and
Observatoire de Haute-Provence (CNRS/OAMP), France
\and
Institut d'Astrophysique et de G\'eophysique de l'Universit\'e de Li\`ege, All\'ee du 6 Ao\^ut 17, B-4000 Li\`ege, Belgium
}
   \date{Received  29 May 2007; accepted  18 September 2007}
  \abstract
   {The theory of stellar evolution can be more closely tested if we have the
   opportunity to measure new quantities. Nowadays, observations of galactic
   RR Lyr stars are available on a time baseline exceeding 100 years. Therefore,
   we can exploit the possibility of investigating period changes, continuing
   the pioneering work started by V.~P.~Tsesevich in 1969.}
   {We collected the available times of maximum
   brightness of the galactic RR Lyr stars in the GEOS RR Lyr database.
   Moreover, we also started new observational
   projects, including surveys with automated telescopes, to characterise the O--C
   diagrams better.}
   {The database we built has proved to be a very powerful tool for tracing the
   period variations through the ages. We analyzed 123 stars showing a clear
   O--C pattern (constant, parabolic or erratic) by means of  different
   least--squares methods.}
   {Clear evidence of period increases or decreases at constant rates has
   been found, suggesting evolutionary effects. The median values are
   $\beta$=+0.14~d~Myr$^{-1}$ for the 27 stars showing a period increase and
   $\beta$=--0.20~d~Myr$^{-1}$ for the 21 stars showing a period decrease. The
   large number of RR Lyr stars showing a period decrease (i.e., blueward
   evolution) is a new and  intriguing result. There is an excess of RR Lyr stars
   showing large, positive $\beta$ values. Moreover, the observed $\beta$
   values are slightly larger than those predicted by theoretical models.}
   {}
\authorrunning{Le Borgne et al.}
\titlerunning{Period changes in RR Lyr stars}
\maketitle
\section{Introduction}
RR Lyr variables are low--mass stars in a core helium burning phase; they fill
the part of the HR diagram where the horizontal branch intersects the
classical instability strip. The crossing of the instability strip
can take place in both directions; as a consequence, the periods
will be either increasing, if the stars evolve from blue to red, or
decreasing, if they evolve from red to blue. Despite its importance
as a test for the stellar evolution theory, the {\it observed} rate of the
period changes is still an unknown quantity. It has been measured in globular
clusters (e.g., Smith \& Sandage \cite{mfifteen}, Lee \cite{lee}, Jurcsik et al.
\cite{omega}), comparing periods determined in different epochs. However, this task
has not been undertaken yet  on the wide population of galactic RR Lyr stars, though
most of them have been studied for tens of years, several of them since the end of
the XIX$^{\rm th}$ century, and therefore the available data are almost continuous.
One of the critical points is to determine the importance of the negative rates,
i.e., the period decreases. Such rates have been observed
in some cluster stars (Jurcsik et  al. \cite{omega}), but it is unclear in which
evolutionary phase they occur (Sweigart \& Renzini \cite{random2}).

In the present paper, we derive the period variation rate of the best observed
RR Lyr stars belonging to the  RRab sub-class. For this purpose we use the
{\it GEOS RR Lyr database} which is described in Sect.~\ref{sect_grrdb}.
The observations of the TAROT telescopes coordinated in the {\it GEOS RR Lyr
Survey} give a strong impulse to this analysis, providing a large number of
times of maximum brightness in the recent years. They complete the effort of
the amateur observers, in particular in the European associations BAV
(Bundesdeutsche Arbeitsgemeinschaft f\"ur Ver\"anderliche Sterne) and GEOS
(Groupe Europ\'een d'Observations Stellaires), which have
been  surveying these stars for years. These new observations are described in
Sect.~\ref{sect_grrs}. Next, Sect.~\ref{sect_datan} shows how the data were analysed.
In Sect.~\ref{sect_cstrate}, we analyse the stars with constant period
variation. Section~\ref{sect_BLZH} is devoted to the study of some
particularities encountered during the present work: Blazhko and
light--time effects. Finally, a discussion of the results is developed in
Sect.~\ref{sect_discuss} in the light of stellar evolution.
\section{The GEOS RR Lyr database and survey: a V.P. Tsesevich's heritage}\label{sect_grrdb}
In his book ``RR Lyrae stars", V. P. Tsesevich (\cite{Tsesevich}) made a summary of the
period variation behavior of RR Lyr stars from the observations made during the
60~years or so after the discovery of these stars. Nowadays, 40 more years have
elapsed, allowing 100~years of data for some stars. It is timely to investigate what
the data accumulated since Tsesevich's work bring to the understanding of the
period variation of RR Lyr stars and of the stellar evolution of horizontal
branch stars.
\subsection{The GEOS RR Lyr Database}
To complete Tsesevich's data collection, the amateur/professional association GEOS
has built a database aiming to put together all possible RR Lyr light maximum times
published in the literature. The publications from the end of the XIX$^{\rm th}$
century up to today were scanned for this purpose. The database continues to be fed
by recent observations from amateur astronomers of the European groups GEOS and BAV.
To date, the database\footnote{The GEOS database is freely accessible on the internet at
the address http://dbrr.ast.obs-mip.fr/. This site is hosted by the Laboratoire
d'Astrophysique de Toulouse-Tarbes, Observatoire Midi-Pyr\'en\'ees, Toulouse, France}
contains about 50000 maximum times from more than 3000 RR Lyr stars.

The stars concerned are galactic RRab and RRc.
The collected times of light maxima were obtained from observations using various
techniques, either visually, with electronic devices, or photographically, and by
hundreds of observers. This heterogeneity is taken into account in the analysis
of the O--C diagrams which are the basis of the present paper.
The O--C diagrams are obtained plotting the differences between the observed and
calculated times of maximum brightness (i.e., the O--C values) vs. the elapsed cycles.
Multicolor photometry of RR Lyr stars (e.g., the extensive $UBV$
light curves obtained by Ol\'ah \& Szeidl \cite{olah} and Szeidl et al. \cite{szeidl})
shows that the colour effects on the determination of the times of maximum brightness
are negligible with respect to the period variations we are searching for, since
they are much smaller than the errors on the determinations themselves.
It is also obvious that the precision of modern observations, either with photoelectric
detectors or CCDs, is much better than visual and photographic observations.
However, the uncertainties of the visual and photographic determinations
are much less than the O--C variations we are studying.
A particularity of photographic observations is that the corresponding published
maximum is often not the result of the analysis of a lightcurve obtained during one
night, but rather one single ``bright" point. These observations were useful
at the time when people searched for the pulsation period of the star, but
unavoidably this method introduces an additional scatter in the O--C diagrams.
\subsection{The TAROT Telescopes}\label{sect_grrs}
The telescopes TAROT ({\it T\'elescope \`a Action Rapide pour les Objets
Transitoires} = {\it Rapid Action Telescope for Transient Objects}; Bringer et al.
\cite{bringer}) are automatic, autonomous observatories composed of 25--cm telescopes
whose first objective is the real-time detection of the optical transient
counterparts of GRBs in the visible (Klotz et al. \cite{klotz})\footnote{TAROT
telescopes were funded by the Centre National de la Recherche Scientifique
(CNRS). The project is led by a team of French laboratories (CESR, LATT and OHP).}.
One of these telescopes is located in the northern hemisphere  at Calern
Observatory (Observatoire de la C\^ote d'Azur, University of Nice, France),
and the second one is in the southern hemisphere at ESO, La Silla, Chile.
The telescopes have a field of view of 1.86$\times$1.86 deg$^2$. During operation
no human action is necessary, neither for calibration and science frames nor for
data processing. The roof of the buildings are retractable to allow rapid movement
to any direction in the sky. TAROT telescopes are able to point in any direction
in less than 5~sec (including CCD operations and filter movements). They can detect
stars of magnitude $V$=16.2 in 10sec exposure time, or $V$=18.2 in 60 sec.
TAROTs are able to monitor a GRB 8~sec after being triggered, about 20~sec after
the gamma ray detection (Bo\"er et al. \cite{Boer}).

A dedicated software program optimises the process: it is able to decide at any time
what is the best observation to do next, given program objects, priorities,
alerts and constraints from the astronomer such as measurement frequency (Bringer et
al., \cite{bringer}). Weather stations are linked to the TAROT computers. If it
rains or if the temperature reaches the dew temperature or the humidity becomes
higher than 90\%, the roof remains closed.
The system performs an automatic data reduction (dark and flat corrections,
astrometric and photometric calibrations) in real time (see Damerdji et al.
\cite{damerdji}). In the case of the images dedicated to the RR Lyr survey,
a catalogue of the measurements is updated after each observation and is
available on the web about 2~min after  the image has been taken.
\subsection{ The GEOS RR Lyr survey with TAROT telescopes}
In addition to the detection and localisation of GRBs, TAROTs follow the CNES
artificial satellites, and a fraction of the observing time is left to other astronomical
programs. Therefore, the efforts made by hundreds of observers over one century
to monitor the maximum times of RR Lyr stars may be continued nowadays by much
more efficient techniques, thanks to the use of these small robotic telescopes.
Indeed, the authors started a monitoring of maxima of RRab stars, called
{\it GEOS RR Lyr survey}, on 2004 January 29 using the northern  TAROT and in
September 2006 using the  southern one.
The capabilities of the system and the time-sharing allow us to schedule the
observation of up to 6 maxima of RR Lyr stars per night and per telescope.
The schedule is sent to the control sofware once per month.
The aim is to obtain at least one maximum per star with $V<12.5$ (Calern
observatory) or with $V<13.5$ (Chile) and per year, in order to check the period
stability on time scales of decades. In practice, the number of maxima observed
for each star of the programme is typically 5--10 per year.
Only stars from the {\it General Catalogue of Variable Stars } (GCVS, Kholopov et al.
\cite{Kholopov}) have been observed so far:
this catalogue contains stars which have often been observed for more than 50 years.
In 3 years of operation, from January 2004 to December 2006, TAROT-Calern
observed 1288 maxima on 130 stars; TAROT-Chile observed 556 maxima on 126 stars
during the first 6 months of operation.

For each maximum, the telescopes obtain an average of 50 measurements in 4.8~hours.
The number of frames having contributed to determine the maximum times is 98000
in three years.
The times of maximum brightness are determined by fitting the data points
by means of a cubic spline function with a non-zero smoothing parameter.
The smoothing parameter was selected case--by--case, depending on the number of
points and their scatter. The error bars are estimated from the data sampling.
The nominal sampling is composed of two consecutive 30--sec exposures taken every
10~min in a time interval of 2 hours, centred around the predicted maximum time and
20~min outside this interval. The resulting mean separation between 2 measurements
is 5.8~min. However, the nominal sampling  may be altered by local events (e.g.,
weather or telescope operation).
For a well covered maximum, the mean error bar is about 0.003~d (4.3 minutes);
usually, error bars range from 0.002 to 0.010~d.
\subsection{ Survey statistics}
In order to check the efficiency of the survey, we analysed the production of
maximum times during the  first three years of TAROT-Calern. The RR Lyr stars in the
GCVS with $V>9.0$ (at minimum) and with $\delta > -10^\circ$
are sorted in bins of 0.5~mag width. RR Lyr itself is not considered, since this
bright variable (7.06--8.12~$V$) is currently monitored with several techniques
(visual, photographic, photoelectric and CCD photometry, medium and high resolution
spectroscopy). Since the magnitudes at minimum come from the GCVS, the photometric
systems are heterogeneous, but close to Johnson $V$ magnitudes.
We also compare the statistics of the TAROT (GEOS)  production (from 2004 to
2006) with those of  the observers of BAV group (from 2002 to 2005).
Both groups published their maximum times in the {\it Information Bulletin on
Variable Stars}: see Le Borgne et al. \cite{LeBorgne04}, \cite{LeBorgne05a},
\cite{LeBorgne05b}, \cite{LeBorgne06a}, \cite{LeBorgne06b} and \cite{LeBorgne07}
for GEOS maxima, and Agerer and H\"ubscher, \cite{Agerer02}, \cite{Agerer03},
H\"ubscher \cite{Hubscher05}, H\"ubscher et al. \cite{Hubscher+05}, \cite{Hubscher06}
for BAV maxima. BAV obtains this result using several telescopes operated by
about 20 observers.

\begin{figure}[]
\begin{center}
\includegraphics[width=0.7\columnwidth, angle=270]{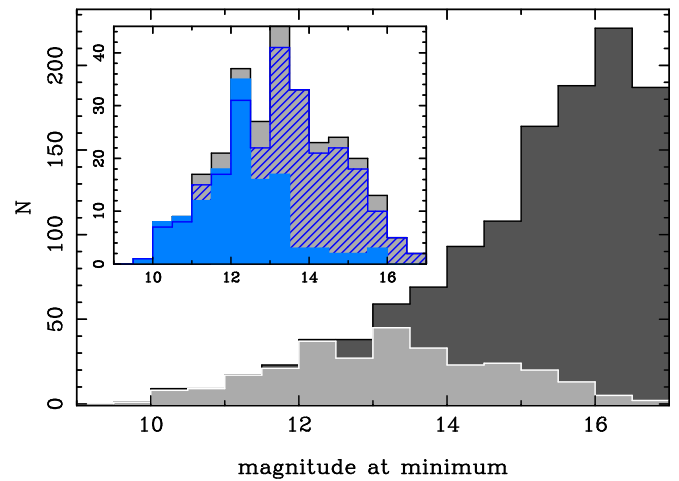}
\caption{\footnotesize Number of RR Lyr stars per magnitude (at minimum) bin
observed from 2004 to 2006.
The observing material shown by the light grey histogram in the main window
of the figure includes the GEOS RR Lyr Survey and the BAV CCD observations
published in the IBVS. The dark grey histogram represents all the RR Lyr stars
included in the GCVS. The enclosed histogram shows details of the contribution of
 the GEOS RR Lyr survey (filled blue) and the BAV (hatched blue).}
\label{stat}
\end{center}
\end{figure}

Table \ref{statT} and Fig.~\ref{stat} illustrate the results of the statistical
analysis. Indeed, the results show that the two groups put together have a completeness
of 97\% up to $V$=12.5 (magnitude at minimum); 93 of the 97 RRab stars listed in
the  GCVS have been observed. Individually, the GEOS and BAV surveys have a completeness
of 88\% and 81\%, respectively. We note that, for the GEOS survey with TAROT
telescopes, $V$=12.5 is a natural limit for a 25--cm telescope. Moreover, both surveys
have a decrease in completeness beyond this magnitude imposed by the typical diameter
of the telescopes (20-30~cm mirror diameter), while a second maximum occurs in the
BAV histogram at magnitude 13.5, corresponding to the use of larger telescopes by
some observers.
\begin{table}
\caption{\label{statT} Number of RR Lyr stars per magnitude (at minimum) bin
observed from 2004 to 2006.}
\centerline{\begin{tabular}{rrrrrr}
\hline
   &  &  &   &   &  \\
  \multicolumn{1}{c}{Magnitude}   &   \multicolumn{5}{c}{Number of stars }    \\
  \multicolumn{1}{c}{at minimum}  &    &   &   &   &  \\
     &  in GCVS   &  \multicolumn{2}{c}{BAV+GEOS} &   \multicolumn{1}{c}{GEOS} & \multicolumn{1}{c}{BAV}   \\
   &  & \multicolumn{2}{c}{2002-2006} & \multicolumn{1}{c}{2004-2006}  & \multicolumn{1}{c}{2002-2005} \\
   &  &  &   &    & \\
\hline
   &  &  &  &  &   \\
 9.5 - 10.0&  1\hspace{0.5cm} &   1\hspace{0.2cm} & 100\% \hspace{0.2cm} &   1\hspace{0.5cm} &   1\hspace{0.5cm} \\
10.0 - 10.5&  9\hspace{0.5cm} &   8\hspace{0.2cm} &  88\% \hspace{0.2cm} &   8\hspace{0.5cm} &   7\hspace{0.5cm} \\
10.5 - 11.0&  9\hspace{0.5cm} &   9\hspace{0.2cm} & 100\% \hspace{0.2cm} &   9\hspace{0.5cm} &   8\hspace{0.5cm} \\
11.0 - 11.5& 17\hspace{0.5cm} &  17\hspace{0.2cm} & 100\% \hspace{0.2cm} &  13\hspace{0.5cm} &  15\hspace{0.5cm} \\
11.5 - 12.0& 23\hspace{0.5cm} &  21\hspace{0.2cm} &  91\% \hspace{0.2cm} &  19\hspace{0.5cm} &  17\hspace{0.5cm} \\
12.0 - 12.5& 38\hspace{0.5cm} &  37\hspace{0.2cm} &  97\% \hspace{0.2cm} &  35\hspace{0.5cm} &  31\hspace{0.5cm} \\
12.5 - 13.0& 38\hspace{0.5cm} &  29\hspace{0.2cm} &  76\% \hspace{0.2cm} &  21\hspace{0.5cm} &  22\hspace{0.5cm} \\
13.0 - 13.5& 59\hspace{0.5cm} &  46\hspace{0.2cm} &  77\% \hspace{0.2cm} &  19\hspace{0.5cm} &  41\hspace{0.5cm} \\
13.5 - 14.0& 69\hspace{0.5cm} &  34\hspace{0.2cm} &  49\% \hspace{0.2cm} &   4\hspace{0.5cm} &  33\hspace{0.5cm} \\
14.0 - 14.5& 93\hspace{0.5cm} &  23\hspace{0.2cm} &  24\% \hspace{0.2cm} &   3\hspace{0.5cm} &  21\hspace{0.5cm} \\
14.5 - 15.0&108\hspace{0.5cm} &  24\hspace{0.2cm} &  22\% \hspace{0.2cm} &   2\hspace{0.5cm} &  22\hspace{0.5cm} \\
15.0 - 15.5&164\hspace{0.5cm} &  20\hspace{0.2cm} &  12\% \hspace{0.2cm} &   2\hspace{0.5cm} &  18\hspace{0.5cm} \\
15.5 - 16.0&188\hspace{0.5cm} &  13\hspace{0.2cm} &   6\% \hspace{0.2cm} &   3\hspace{0.5cm} &  10\hspace{0.5cm} \\
16.0 - 16.5&222\hspace{0.5cm} &   5\hspace{0.2cm} &   2\% \hspace{0.2cm} &   0\hspace{0.5cm} &   5\hspace{0.5cm} \\
16.5 - 17.0&187\hspace{0.5cm} &   2\hspace{0.2cm} &   1\% \hspace{0.2cm} &   0\hspace{0.5cm} &   2\hspace{0.5cm} \\
   &  &  &  &    &    \\
\hline
\end{tabular}}
\end{table}
\begin{table}
\caption{Inventory of the O--C patterns in the RRab stars of
our sample}
\begin{tabular}{lrl}
\hline
\noalign{\smallskip}
\multicolumn{1}{c}{O--C pattern} & \multicolumn{1}{c}{N}&\multicolumn{1}{c}{Representative cases}  \\
\noalign{\smallskip}
\hline
\noalign{\smallskip}
Constant period & 54 & \\
Increasing period & 27 & \object{BN Aqr}, \object{RR Leo}\\
Decreasing period & 21 & \object{SW And}, \object{AH Cam}, \object{V964 Ori} \\
Irregular slow variations & 17 & \object{UY Boo},  \object{RW Dra}\\
Single abrupt change & 4 & \object{SZ Hya} \\
\noalign{\smallskip}
Total & 123 \\
\noalign{\smallskip}
\hline
\label{inven}
\end{tabular}
\end{table}
\begin{table*}
\caption{Refined linear elements for stars showing a constant period.
The note $^B$  beside the value of the standard deviation indicates
the stars having a known Blazhko effect.}
\begin{tabular}{ ll rr rr  l}
\hline
\noalign{\smallskip}
 \multicolumn{2}{c}{Star} & \multicolumn{1}{c}{N$_{\rm max}$}&
\multicolumn{1}{c}{Time coverage} &
 \multicolumn{1}{c}{Epoch}&
\multicolumn{1}{c}{Period}&
\multicolumn{1}{c}{s.d.} \\
 \multicolumn{2}{c}{} & \multicolumn{1}{c}{}&
\multicolumn{1}{c}{[years]} &
 \multicolumn{1}{c}{[HJD -- 2400000]} &
\multicolumn{1}{c}{[d]}&
\multicolumn{1}{c}{[d]} \\
\noalign{\smallskip}
\hline
\noalign{\smallskip}
AT & And     &  66 & 101 (1906-2007) & 36109.5000 $\pm$ 0.0030 & 0.61691445 $\pm$ 0.00000014 & 0.0169 \\
DE & And     &  29 &  64 (1935-1999) & 39764.5234 $\pm$ 0.0033 & 0.45363536 $\pm$ 0.00000019 & 0.0176 \\
GV & And     &  54 &  62 (1937-1999) & 40171.3320 $\pm$ 0.0050 & 0.52809310 $\pm$ 0.00000033 & 0.0319 \\
OV & And     & 100 &  75 (1929-2004) & 39469.2891 $\pm$ 0.0012 & 0.47058114 $\pm$ 0.00000007 & 0.0111 \\
TZ & Aqr     &  34 &  92 (1915-2007) & 37548.9023 $\pm$ 0.0022 & 0.57119429 $\pm$ 0.00000010 & 0.0128 \\
YZ & Aqr     &  51 & 112 (1894-2006) & 33343.2617 $\pm$ 0.0033 & 0.55193269 $\pm$ 0.00000017 & 0.0218 \\
DN & Aqr     &  35 &  71 (1936-2007) & 40915.3125 $\pm$ 0.0050 & 0.63375407 $\pm$ 0.00000039 & 0.0268 \\
TZ & Aur     & 140 &  94 (1913-2007) & 37946.4922 $\pm$ 0.0006 & 0.39167482 $\pm$ 0.00000002 & 0.0051 \\
BH & Aur     &  41 & 100 (1906-2006) & 37352.4727 $\pm$ 0.0024 & 0.45608902 $\pm$ 0.00000007 & 0.0086 \\
SS & Cnc     & 104 &  96 (1909-2005) & 36201.3906 $\pm$ 0.0012 & 0.36733848 $\pm$ 0.00000003 & 0.0117 \\
RX & CVn     &  56 &  98 (1907-2005) & 36904.2266 $\pm$ 0.0024 & 0.54002589 $\pm$ 0.00000012 & 0.0176 \\
AL & CMi     &  41 &  82 (1925-2007) & 36607.3047 $\pm$ 0.0046 & 0.55051273 $\pm$ 0.00000020 & 0.0243 \\
UU & Cet     &  30 &  82 (1917-1999) & 37582.4336 $\pm$ 0.0077 & 0.60607427 $\pm$ 0.00000056 & 0.0380$^B$ \\
SU & Col     &  25 &  65 (1937-2002) & 39410.5234 $\pm$ 0.0064 & 0.48735806 $\pm$ 0.00000035 & 0.0302 \\
ST & Com     &  62 &  90 (1916-2006) & 37652.6680 $\pm$ 0.0022 & 0.59892809 $\pm$ 0.00000011 & 0.0163 \\
W & Crt      &  23 &  70 (1935-2005) & 39910.8633 $\pm$ 0.0014 & 0.41201425 $\pm$ 0.00000007 & 0.0063 \\
UY & Cyg     & 106 & 105 (1901-2006) & 34862.3203 $\pm$ 0.0014 & 0.56070578 $\pm$ 0.00000006 & 0.0116 \\
BV & Del     &  28 &  77 (1928-2005) & 39354.2539 $\pm$ 0.0037 & 0.42345089 $\pm$ 0.00000015 & 0.0184 \\
BT & Dra     &  63 & 101 (1905-2006) & 35868.4648 $\pm$ 0.0028 & 0.58867335 $\pm$ 0.00000014 & 0.0205 \\
SZ & Gem     &  74 &  85 (1922-2007) & 38360.8438 $\pm$ 0.0015 & 0.50113541 $\pm$ 0.00000007 & 0.0108 \\
GI & Gem     &  63 &  66 (1941-2007) & 36971.7148 $\pm$ 0.0016 & 0.43326673 $\pm$ 0.00000007 & 0.0122 \\
TW & Her     & 226 &  94 (1912-2006) & 36784.3828 $\pm$ 0.0004 & 0.39960003 $\pm$ 0.00000002 & 0.0062 \\
BD & Her     &  62 &  97 (1900-1997) & 33064.4766 $\pm$ 0.0043 & 0.47390789 $\pm$ 0.00000025 & 0.0317 \\
EE & Her     &  33 &  57 (1936-1993) & 38853.6250 $\pm$ 0.0026 & 0.49553666 $\pm$ 0.00000025 & 0.0129 \\
V524 & Her   & 122 &  56 (1930-1986) & 36318.5156 $\pm$ 0.0032 & 0.48186663 $\pm$ 0.00000022 & 0.0326 \\
GO & Hya     &  39 &  78 (1929-2007) & 35821.6641 $\pm$ 0.0052 & 0.63643533 $\pm$ 0.00000028 & 0.0317 \\
RX & Leo     &  23 &  72 (1935-2007) & 38431.9336 $\pm$ 0.0055 & 0.65341568 $\pm$ 0.00000027 & 0.0115 \\
ST & Leo     &  96 &  80 (1926-2006) & 39260.4219 $\pm$ 0.0007 & 0.47798404 $\pm$ 0.00000004 & 0.0068 \\
WW & Leo     &  40 &  72 (1935-2007) & 39801.6211 $\pm$ 0.0031 & 0.60284597 $\pm$ 0.00000021 & 0.0197 \\
AA & Leo     &  22 &  52 (1953-2005) & 47271.3828 $\pm$ 0.0021 & 0.59865355 $\pm$ 0.00000023 & 0.0088 \\
AX & Leo     &  35 &  79 (1927-2006) & 37026.5742 $\pm$ 0.0041 & 0.72682691 $\pm$ 0.00000023 & 0.0230 \\
V & LMi      &  34 &  95 (1912-2007) & 36630.6094 $\pm$ 0.0018 & 0.54391927 $\pm$ 0.00000006 & 0.0065 \\
VY & Lib     &  21 &  77 (1914-1991) & 33773.3203 $\pm$ 0.0035 & 0.53394097 $\pm$ 0.00000028 & 0.0162 \\
TT & Lyn     &  50 &  64 (1943-2007) & 43580.4023 $\pm$ 0.0032 & 0.59743237 $\pm$ 0.00000023 & 0.0224 \\
TW & Lyn     &  70 &  48 (1956-2004) & 45022.4961 $\pm$ 0.0022 & 0.48186052 $\pm$ 0.00000013 & 0.0187 \\
CN & Lyr     &  72 & 105 (1901-2006) & 36079.3242 $\pm$ 0.0018 & 0.41138276 $\pm$ 0.00000005 & 0.0104 \\
IO & Lyr     &  95 &  97 (1909-2006) & 36276.2305 $\pm$ 0.0013 & 0.57712215 $\pm$ 0.00000007 & 0.0113 \\
KX & Lyr     &  62 &  65 (1940-2005) & 42019.2461 $\pm$ 0.0036 & 0.44090429 $\pm$ 0.00000017 & 0.0188 \\
LX & Lyr     &  33 &  64 (1940-2004) & 36287.2930 $\pm$ 0.0026 & 0.54548401 $\pm$ 0.00000022 & 0.0148 \\
RV & Oct     & 140 &  91 (1900-1991) & 31874.5703 $\pm$ 0.0029 & 0.57116264 $\pm$ 0.00000022 & 0.0347 \\
ST & Oph     &  29 &  98 (1908-2006) & 36079.3438 $\pm$ 0.0028 & 0.45035613 $\pm$ 0.00000011 & 0.0144 \\
V531 & Oph   &  25 &  64 (1928-1992) & 39618.5156 $\pm$ 0.0033 & 0.55365324 $\pm$ 0.00000022 & 0.0166 \\
V2033 & Oph  &  24 &  52 (1940-1992) & 39270.5352 $\pm$ 0.0047 & 0.56582946 $\pm$ 0.00000043 & 0.0201 \\
AO & Peg     &  23 &  80 (1925-2005) & 37909.3984 $\pm$ 0.0021 & 0.54724342 $\pm$ 0.00000010 & 0.0092 \\
BF & Peg     &  24 &  73 (1931-2004) & 39369.1758 $\pm$ 0.0062 & 0.49580538 $\pm$ 0.00000029 & 0.0301 \\
CG & Peg     & 109 &  51 (1956-2007) & 45576.4414 $\pm$ 0.0015 & 0.46713728 $\pm$ 0.00000010 & 0.0144 \\
DZ & Peg     &  63 &  97 (1910-2007) & 36511.4062 $\pm$ 0.0043 & 0.60734892 $\pm$ 0.00000026 & 0.0313 \\
VY & Ser     &  41 &  74 (1932-2006) & 39615.9297 $\pm$ 0.0049 & 0.71409613 $\pm$ 0.00000038 & 0.0225 \\
AN & Ser     & 106 & 107 (1899-2006) & 34265.2461 $\pm$ 0.0014 & 0.52207130 $\pm$ 0.00000007 & 0.0131$^B$ \\
DF & Ser     &  42 & 105 (1900-2005) & 34195.6641 $\pm$ 0.0019 & 0.43779647 $\pm$ 0.00000008 & 0.0125 \\
UU & Vir     &  23 &  97 (1903-2000) & 34509.2617 $\pm$ 0.0012 & 0.47560632 $\pm$ 0.00000005 & 0.0059 \\
WY & Vir     &  37 &  87 (1914-2001) & 38060.6328 $\pm$ 0.0068 & 0.60935342 $\pm$ 0.00000058 & 0.0409$^B$ \\
AS & Vir     &  19 &  84 (1916-2000) & 37052.4844 $\pm$ 0.0123 & 0.55342495 $\pm$ 0.00000075 & 0.0407 \\
AV & Vir     &  25 &  93 (1913-2006) & 37016.4492 $\pm$ 0.0025 & 0.65690929 $\pm$ 0.00000012 & 0.0097 \\
\noalign{\smallskip}
\hline
\label{cst}
\end{tabular}
\end{table*}
\begin{table*}
\caption{Parabolic elements for RRab stars showing a well--defined
linearly increasing period. The note $^B$ beside the value of the standard deviation
indicates the stars having  a known Blazhko effect and discussed in the text.}
\begin{tabular}{rr c rrr l  r  rr}
\hline
\noalign{\smallskip}
 \multicolumn{1}{c}{Star} & \multicolumn{1}{c}{N$_{\rm max}$}&
\multicolumn{1}{c}{Time coverage} &
 \multicolumn{1}{c}{Epoch}& \multicolumn{1}{c}{Period}&
 \multicolumn{1}{c}{Quad. term}&
\multicolumn{1}{c}{s.d.} &
\multicolumn{1}{c}{dP/dt}&
\multicolumn{1}{c}{$\beta$}&
\multicolumn{1}{c}{$\alpha$} \\
 \multicolumn{1}{c}{} & \multicolumn{1}{c}{}&
\multicolumn{1}{c}{[years]} &
 \multicolumn{1}{c}{[HJD-2400000]}& \multicolumn{1}{c}{[d]}&
 \multicolumn{1}{c}{[10$^{-10}$ $\cdot$d]}&
\multicolumn{1}{c}{[d]} &
\multicolumn{1}{c}{[10$^{-10}$ $\cdot$d/d]}&
\multicolumn{1}{c}{[d~Myr$^{-1}$]}&
\multicolumn{1}{c}{}[Myr$^{-1}$] \\
\noalign{\smallskip}
\hline
\noalign{\smallskip}
XX And & 182 & 105           &   54106.3508 &       0.72276008 &       1.427 & 0.0167 &   3.95 & 0.144 & 0.200 \\
       &     &  (1902-2007)  &$\pm$  0.0027 & $\pm$ 0.00000029 & $\pm$ 0.066 &        &$\pm$ 0.18 & $\pm$0.007 & $\pm$0.009\\
\noalign{\smallskip}
BN Aqr &  78 & 110  &            53979.4964 &       0.46967415 &       3.173 & 0.0149 &     13.51 & 0.494 & 1.051 \\
       &     &   (1896-2006) &$\pm$  0.0049 & $\pm$ 0.00000026 & $\pm$ 0.030 &        &$\pm$ 0.13 & $\pm$0.005 & $\pm$0.010 \\
\noalign{\smallskip}
V341 Aql & 117 & 94  &           53999.4346 &        0.57802322 &    0.613 & 0.0130 &       2.12 & 0.078 & 0.134 \\
       &     &   (1912-2006) &$\pm$  0.0025 & $\pm$ 0.00000027 & $\pm$ 0.056 &        &$\pm$ 0.19 &$\pm$0.007 & $\pm$0.012 \\
\noalign{\smallskip}
 X Ari &  112 &  93  &           54107.2779 &       0.65116811 &       4.763 & 0.0135 &     14.63 & 0.535 & 0.821 \\
       &     &   (1914-2007) &$\pm$  0.0021 & $\pm$ 0.00000028 & $\pm$ 0.060 &       & $\pm$ 0.18 & $\pm$0.007 & $\pm$0.010 \\
\noalign{\smallskip}
RS Boo & 325 & 107  &            54113.6372 &       0.37734035 &       0.268 & 0.0093$^B$ &     1.42 & 0.052 & 0.137\\
       &     &   (1900-2007) &$\pm$  0.0011 & $\pm$ 0.00000006 & $\pm$ 0.007 &       & $\pm$ 0.04 & $\pm$0.001 & $\pm$0.004 \\
\noalign{\smallskip}
SW Boo & 52  &  98  &            53919.4937 &       0.51355094 &       4.639 & 0.0080$^B$ &     18.07 & 0.660 & 1.286\\
       &     &   (1908-2006) &$\pm$  0.0019 & $\pm$ 0.00000017 & $\pm$ 0.030 &        & $\pm$ 0.12 & $\pm$0.004 & $\pm$0.008\\
\noalign{\smallskip}
UU Boo & 88  &  99  &            53904.4944 &       0.45693375 &       2.231 & 0.0116 &      9.77 & 0.357 & 0.781\\
       &     &   (1907-2006) &$\pm$  0.0023 & $\pm$ 0.00000017 & $\pm$ 0.025 &        & $\pm$ 0.11 &$\pm$0.004 & $\pm$0.009 \\
\noalign{\smallskip}
RW Cnc & 95  &  90  &            53746.5042 &       0.54721602 &       1.937 & 0.0169 &      7.08 & 0.259 & 0.473\\
       &     &   (1916-2006) &$\pm$  0.0035 & $\pm$ 0.00000033 & $\pm$ 0.054 &        & $\pm$ 0.20 &$\pm$0.007 & $\pm$0.013 \\
\noalign{\smallskip}
TT Cnc & 97  &  91  &            54112.4014 &       0.56345656 &       1.229 & 0.0132$^B$ &      4.36 & 0.159 & 0.283\\
       &     &   (1916-2007) &$\pm$  0.0023 & $\pm$ 0.00000029 & $\pm$ 0.058 &        & $\pm$ 0.21 & $\pm$0.008 & $\pm$0.013\\
\noalign{\smallskip}
AN Cnc & 29  &  96  &            53752.5259 &       0.54316354 &       0.555 & 0.0122 &      2.04 & 0.075 & 0.137\\
       &     &  (1910-2006)  &$\pm$  0.0043 & $\pm$ 0.00000037 & $\pm$ 0.067 &        & $\pm$ 0.25 & $\pm$0.009 & $\pm$0.017\\
\noalign{\smallskip}
SW CVn & 19  &  95  &            53478.5374 &       0.44167101 &       2.033 & 0.0066 &      9.21 & 0.336 & 0.762\\
       &     &   (1910-2005) &$\pm$  0.0023 & $\pm$ 0.00000024 & $\pm$ 0.036 &        & $\pm$ 0.16 & $\pm$0.006 & $\pm$0.013\\
\noalign{\smallskip}
EZ Cep & 89  &  105  &           53750.3391 &      0.37900204 &       0.296 & 0.0236 &      1.56 & 0.057 & 0.151 \\
       &     &   (1901-2006) &$\pm$  0.0067 & $\pm$ 0.00000027 & $\pm$ 0.028 &        & $\pm$ 0.15 & $\pm$0.005 & $\pm$0.014 \\
\noalign{\smallskip}
RR Cet & 113  & 100  &           54090.2963 &       0.55302909 &      0.231 & 0.0075 &     0.84 & 0.031 & 0.055 \\
       &     &   (1906-2006) &$\pm$  0.0014 & $\pm$ 0.00000011 & $\pm$ 0.020 &        & $\pm$ 0.07 &$\pm$ 0.003 &$\pm$ 0.005 \\
\noalign{\smallskip}
TV CrB & 124 &  104 &            53903.4827 &       0.58461776 &      0.893 & 0.0311 &        3.06 & 0.112 & 0.191\\
       &     &    (1902-2006)&$\pm$  0.0069 & $\pm$ 0.00000047 & $\pm$ 0.074  &        & $\pm$ 0.25 &$\pm$0.009 & $\pm$0.016\\
\noalign{\smallskip}
DM Cyg  & 223&  106 &            54035.4065 &      0.41986367 &       0.523 & 0.0066$^B$ &      2.49 & 0.091 & 0.217\\
       &     &    (1900-2006)&$\pm$  0.0009 & $\pm$ 0.00000007 & $\pm$ 0.010  &        & $\pm$ 0.05 & $\pm$0.002 & $\pm$0.004 \\
\noalign{\smallskip}
V684 Cyg  & 36&  53 &            44253.2522 &      0.54099474 &       5.222 & 0.0227 &      19.30 & 0.705 & 1.304\\
       &     &    (1927-1980)&$\pm$  0.0084 & $\pm$ 0.00000119 & $\pm$ 0.314  &        & $\pm$ 1.16 & $\pm$0.042 &$\pm$0.079 \\
\noalign{\smallskip}
SU Dra  & 189&  103 &            54111.5136 &      0.66042294 &       0.561 & 0.0124 &       1.70 & 0.062 & 0.094\\
       &     &    (1904-2007)&$\pm$  0.0020 & $\pm$ 0.00000019 & $\pm$ 0.034  &        & $\pm$ 0.10 & $\pm$0.004 & $\pm$0.006\\
\noalign{\smallskip}
SV Eri  & 54&  104 &             53998.8128 &      0.71387846 &       20.72 & 0.0429 &      58.06 & 2.121 & 2.972\\
       &     &   (1904-2006) &$\pm$  0.0153 & $\pm$ 0.00000142 & $\pm$ 0.304  &        & $\pm$ 0.85 & $\pm$0.031 & $\pm$0.044\\
\noalign{\smallskip}
VZ Her& 203 &  107 &             53967.4819 &       0.44033227 &       0.678 & 0.0207 &     3.08 & 0.112 & 0.255\\
       &     &    (1899-2006)&$\pm$  0.0031 & $\pm$ 0.00000018 & $\pm$ 0.023  &        & $\pm$ 0.10 &$\pm$0.004 & $\pm$0.009 \\
\noalign{\smallskip}
RR Leo& 298 &  109 &             54124.4177 &      0.45240129 &        1.869 & 0.0072 &      8.26 & 0.302 & 0.668\\
       &     &   (1898-2007) &$\pm$  0.0009 & $\pm$ 0.00000006 & $\pm$ 0.009  &        & $\pm$ 0.04 & $\pm$0.001 & $\pm$0.003\\
\noalign{\smallskip}
TV Leo& 27 &  89 &               52373.1085 &      0.67286445 &        3.271 & 0.0307 &      9.72 & 0.355 & 0.528\\
       &     &  (1913-2002)  &$\pm$  0.0184 & $\pm$ 0.00000147 & $\pm$ 0.284  &        & $\pm$ 0.84 & $\pm$0.031 & $\pm$0.046\\
\noalign{\smallskip}
U Lep& 33 &  118 &               54114.6303 &       0.58147954 &       0.685 & 0.0077 &      2.36 & 0.086 & 0.148\\
       &     &  (1899-2007)  &$\pm$  0.0019 & $\pm$ 0.00000018 & $\pm$ 0.030  &        & $\pm$ 0.10 & $\pm$0.004 & $\pm$0.006\\
\noalign{\smallskip}
AV Peg& 443 &  103 &             54060.3926 &       0.39038092 &       0.894 & 0.0084 &     4.58 & 0.167 & 0.429\\
       &     &  (1903-2006)  &$\pm$  0.0010 & $\pm$ 0.00000006 & $\pm$ 0.010  &        & $\pm$ 0.05 & $\pm$0.002 & $\pm$0.005 \\
\noalign{\smallskip}
AR Per& 153 &  107 &             54124.4299 &       0.42555066 &       0.136 & 0.0057 &     0.64 & 0.023 & 0.055\\
     &     &  (1900-2007)  &$\pm$  0.0008 & $\pm$ 0.00000006 & $\pm$ 0.009  &        & $\pm$ 0.04 &  $\pm$0.001 & $\pm$0.003\\
\noalign{\smallskip}
RY Psc& 42 &  96 &               54037.5428 &       0.52973431 &       0.863 & 0.0205 &     3.26 & 0.119 & 0.225\\
       &     &  (1912-2006)  &$\pm$  0.0086 & $\pm$ 0.00000062 & $\pm$ 0.098  &        & $\pm$ 0.37 & $\pm$0.014 & $\pm$0.026 \\
\noalign{\smallskip}
CS Ser& 27 &  102 &               53530.4879 &       0.52679833 &       0.747 & 0.0142 &     2.84 & 0.104 & 0.197\\
      &     &  (1903-2005)  &$\pm$    0.0090 & $\pm$ 0.00000050 & $\pm$ 0.066  &        & $\pm$ 0.25 & $\pm$0.009 & $\pm$0.017 \\
\noalign{\smallskip}
BB Vir& 25 &  94  &               53849.6121 &       0.47110683 &       1.083 & 0.0193 &     4.60 & 0.168 & 0.357\\
      &     &  (1903-2005)  &$\pm$    0.0062 & $\pm$ 0.00000060 & $\pm$ 0.089  &        & $\pm$ 0.38 & $\pm$0.014 & $\pm$0.029 \\
\noalign{\smallskip}
\hline
\end{tabular}
\label{tinc}
\end{table*}
\begin{figure*}[h]
\begin{center}
\includegraphics[width=1.95\columnwidth]{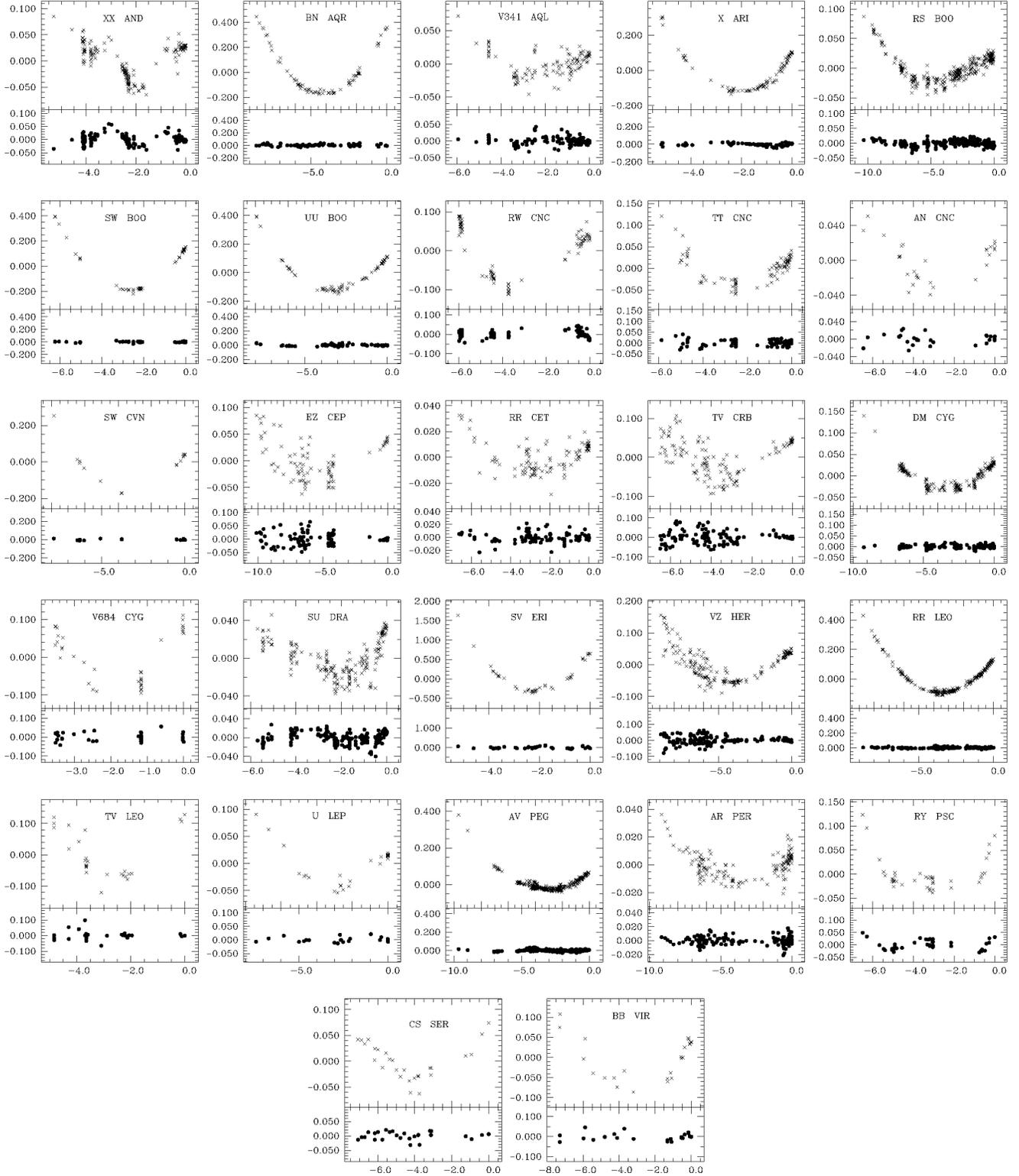}
\caption{\footnotesize O--C values for RRab stars. Filled circles indicate O--C
values calculated from the parabolic fit (period increasing at a constant rate),
crosses from the linear fit (constant period). On the $x$--axis we plot the
elapsed cycles (E/10000); on the ordinate one we plot the  O--C values in days.
The cases of \object{V341 Aql}, \object{RS Boo}, \object{RW Cnc}, \object{EZ Cep} and \object{SU Dra}  are discussed in
the text.}
\label{incf}
\end{center}
\end{figure*}
\begin{table*}
\caption{Parabolic elements for RRab stars showing a well--defined
linearly decreasing period. The note $^B$ beside the value of the standard deviation
indicates the stars having  a known Blazhko effect and discussed in the text.}
\begin{tabular}{rr c rrr l  r  rr}
\hline
\noalign{\smallskip}
 \multicolumn{1}{c}{Star} & \multicolumn{1}{c}{N$_{\rm max}$}&
\multicolumn{1}{c}{Time coverage} &
 \multicolumn{1}{c}{Epoch}& \multicolumn{1}{c}{Period}&
 \multicolumn{1}{c}{Quad. term}&
\multicolumn{1}{c}{s.d.} &
\multicolumn{1}{c}{dP/dt}&
\multicolumn{1}{c}{$\beta$}&
\multicolumn{1}{c}{$\alpha$}
\\
 \multicolumn{1}{c}{} & \multicolumn{1}{c}{}&
\multicolumn{1}{c}{[years]} &
 \multicolumn{1}{c}{[HJD-2400000]}& \multicolumn{1}{c}{[d]}&
\multicolumn{1}{c}{[10$^{-10}$ $\cdot$d/d]}&
\multicolumn{1}{c}{[d]} &
\multicolumn{1}{c}{[10$^{-10}$ $\cdot$d/d]}&
\multicolumn{1}{c}{[d~Myr$^{-1}$]}&
\multicolumn{1}{c}{[Myr$^{-1}$]} \\
\noalign{\smallskip}
\hline
\noalign{\smallskip}
\noalign{\smallskip}
SW And & 393 & 111  &         54093.3336 &       0.44226187 &    --~1.013 & 0.0132$^B$ &   --~4.58 & --~0.167 & --~0.378\\
       &     & (1894-2005)&$\pm$  0.0016 & $\pm$ 0.00000010 & $\pm$ 0.013 &        &$\pm$ 0.06 & $\pm$0.002 & $\pm$0.005\\
\noalign{\smallskip}
SX Aqr &  97 & 93  &          54018.3651 &        0.53570932 &    --~0.537 & 0.0139 &   --~2.00 & --~0.073 & --~0.137\\
       &     & (1913-2006)&$\pm$  0.0031 & $\pm$ 0.00000025 & $\pm$ 0.040 &        &$\pm$ 0.15 & $\pm$0.005 & $\pm$0.010 \\
\noalign{\smallskip}
BR Aqr &  66 & 105  &         54048.3622 &       0.48187046&    --~0.950 & 0.0128 &   --~3.94 & --~0.144 & --~0.299\\
       &     & (1901-2006)&$\pm$  0.0045 & $\pm$ 0.00000024 & $\pm$ 0.031 &        &$\pm$ 0.13 &$\pm$0.005 & $\pm$0.010  \\
\noalign{\smallskip}
CP Aqr &  82 & 72  &          54024.3267 &       0.46340227&    --~0.516 & 0.0103 &   --~2.23 & --~0.081 & --~0.176\\
       &     &(1912-2006)& $\pm$  0.0025 & $\pm$ 0.00000017 & $\pm$ 0.030 &        &$\pm$ 0.13 &$\pm$0.005 & $\pm$0.010 \\
\noalign{\smallskip}
TW Boo & 92  &  92  &         53918.4570 &       0.53226977 &    --~0.503 & 0.0095 &   --~1.89 & --0.069 & --~0.130\\
       &     & (1914-2006)&$\pm$  0.0015 & $\pm$ 0.00000019 & $\pm$ 0.036 &        & $\pm$ 0.14 &$\pm$0.005 & $\pm$0.009  \\
\noalign{\smallskip}
AH Cam & 109  &  56  &        54119.3136 &       0.36871531 &   --~2.397 & 0.0167$^B$ &    --~13.00 & --~0.475 & --~1.288\\
       &     & (1951-2007)&$\pm$  0.0029 & $\pm$ 0.00000039 & $\pm$ 0.076 &        & $\pm$ 0.41 & $\pm$0.015 & $\pm$0.041\\
\noalign{\smallskip}
AP Cnc & 24  &  72  &         52635.7481 &       0.53293273 &   --~2.178 & 0.0259 &    --~8.17 & --~0.299 & --~0.560 \\
       &     & (1930-2002)&$\pm$  0.0192 & $\pm$ 0.00000138 & $\pm$ 0.249 &        & $\pm$ 0.93 & $\pm$0.034 & $\pm$0.064\\
\noalign{\smallskip}
W CVn & 110  &  105  &        54121.6211 &       0.55175472 &   --~0.412 & 0.0073 &    --~1.49 & --~0.055 & --~0.099\\
       &     & (1902-2006)&$\pm$  0.0012 & $\pm$ 0.00000014 & $\pm$ 0.024 &        & $\pm$ 0.09 &$\pm$0.003 & $\pm$0.006 \\
\noalign{\smallskip}
RV Cap & 258  &  95 &         51982.7919 &       0.44773748 &   --~1.260 & 0.0270$^B$ &    --~5.63 & --~0.206 & --~0.459\\
       &     & (1906-2001)&$\pm$  0.0144 & $\pm$ 0.00000069 & $\pm$ 0.084 &        & $\pm$ 0.37 & $\pm$0.014 & $\pm$0.030\\
\noalign{\smallskip}
RX Cet & 120  &  111 &        52172.1923 &       0.57368560 &   --~1.950 & 0.0280 &    --~6.80 & --~0.248 & --~0.433\\
       &     & (1890-2001)&$\pm$  0.0128 & $\pm$ 0.00000061 & $\pm$ 0.075 &        & $\pm$ 0.26 &$\pm$0.010 & $\pm$0.017 \\
\noalign{\smallskip}
RZ Cet &  39  &  77  &        54036.5318 &       0.51059118 &   --~4.032 & 0.0336 &    --~15.79 &--~0.577 & --~1.130\\
       &     & (1929-2006)&$\pm$  0.0133 & $\pm$ 0.00000127 & $\pm$ 0.224 &        & $\pm$ 0.88 & $\pm$0.032 & $\pm$0.063\\
\noalign{\smallskip}
S Com  &  89 &  96 &          54118.5568 &       0.58658461 &    --~0.773 & 0.0097 &   --~2.64 & --~0.096 & --~0.164\\
       &     & (1911-2007)&$\pm$  0.0019 & $\pm$ 0.00000018 & $\pm$ 0.032  &        & $\pm$ 0.11 &$\pm$0.004 & $\pm$0.007 \\
\noalign{\smallskip}
VX Her& 164 &  90 &           53919.4514 &        0.45536088 &    --~0.922 & 0.0107 &   --~4.05 & --~0.148 & --~0.325\\
       &     & (1916-2006)&$\pm$  0.0016 & $\pm$ 0.00000013 & $\pm$ 0.022  &        & $\pm$ 0.09 & $\pm$0.003 & $\pm$0.008\\
\noalign{\smallskip}
V394 Her& 105 &  77 &         53932.4381 &      0.43605116 &    --~0.748 & 0.0205 &   --~3.43 &  --~0.125 & --~0.287 \\
       &     & (1929-2001)&$\pm$  0.0049 & $\pm$ 0.00000042 & $\pm$ 0.070  &        & $\pm$ 0.32 &$\pm$0.012 & $\pm$0.027  \\
\noalign{\smallskip}
RZ Lyr& 396 &  111 &          53983.3800 &       0.51123785 &    --~1.312 & 0.0172$^B$ &   --~5.13 & --~0.188 & --~0.367\\
       &     &(1895-2006) &$\pm$  0.0024 & $\pm$ 0.00000014 & $\pm$ 0.019  &        & $\pm$ 0.08 & $\pm$0.005 & $\pm$0.003 \\
\noalign{\smallskip}
V1095 Oph& 32 &  64 &         48832.4223 &       0.45877735 &    --~1.246 & 0.0212 &   --~5.43 &  --~0.198 & --~0.433 \\
       &    &(1928-1992)  &$\pm$  0.0074 & $\pm$  0.00000091 & $\pm$ 0.174  &        & $\pm$ 0.76 & $\pm$0.028 & $\pm$0.060 \\
\noalign{\smallskip}
V964 Ori& 24 &  104 &         54080.5994 &       0.50463562 &    --~2.730 & 0.0087 &   --~10.82 & --~0.395 & --~0.783\\
       &     & (1902-2006)&$\pm$  0.0045 & $\pm$ 0.00000028 & $\pm$ 0.039  &        & $\pm$ 0.15 & $\pm$0.006 & $\pm$0.011\\
\noalign{\smallskip}
BH Peg& 131 &  73 &           54059.2925 &        0.64098176 &    --~2.785 & 0.0168$^B$ &    --~8.69  & --~0.317 & --~0.495\\
       &    & (1932-2006) &$\pm$  0.0028 & $\pm$ 0.00000035 & $\pm$ 0.084  &        & $\pm$ 0.26 & $\pm$0.010 & $\pm$0.015 \\
\noalign{\smallskip}
SW Psc& 30 &  54 &            39410.3459 &        0.52124757 &    --~5.395 & 0.0272$^B$ &    --~20.70  & --~0.756 & --~1.451\\
       &     & (1912-1966)&$\pm$  0.0079 & $\pm$ 0.00000152 & $\pm$ 0.433  &        & $\pm$ 1.66 &  $\pm$0.061 &  $\pm$0.116\\
\noalign{\smallskip}
HK Pup& 38 &  76 &            54100.6304 &        0.73421141 &    --~7.216 & 0.0285 &    --~19.66  & --~0.718 & --~0.978\\
       &     & (1930-2006)&$\pm$  0.0161 & $\pm$ 0.00000165 & $\pm$ 0.412  &        & $\pm$ 1.12 & $\pm$0.041 & $\pm$0.056\\
\noalign{\smallskip}
AT Vir& 25 &  89 &            53487.4580 &        0.52577511&    --~3.350 & 0.0140 &    --~12.74  & --~0.466 & --~0.885\\
       &     & (1915-2004)&$\pm$  0.0049 & $\pm$ 0.00000047 & $\pm$ 0.086  &        & $\pm$ 0.33 &  $\pm$0.012 &  $\pm$0.023\\
\hline
\end{tabular}
\label{tdec}
\end{table*}
\begin{figure*}
\begin{center}
\includegraphics[width=1.95\columnwidth]{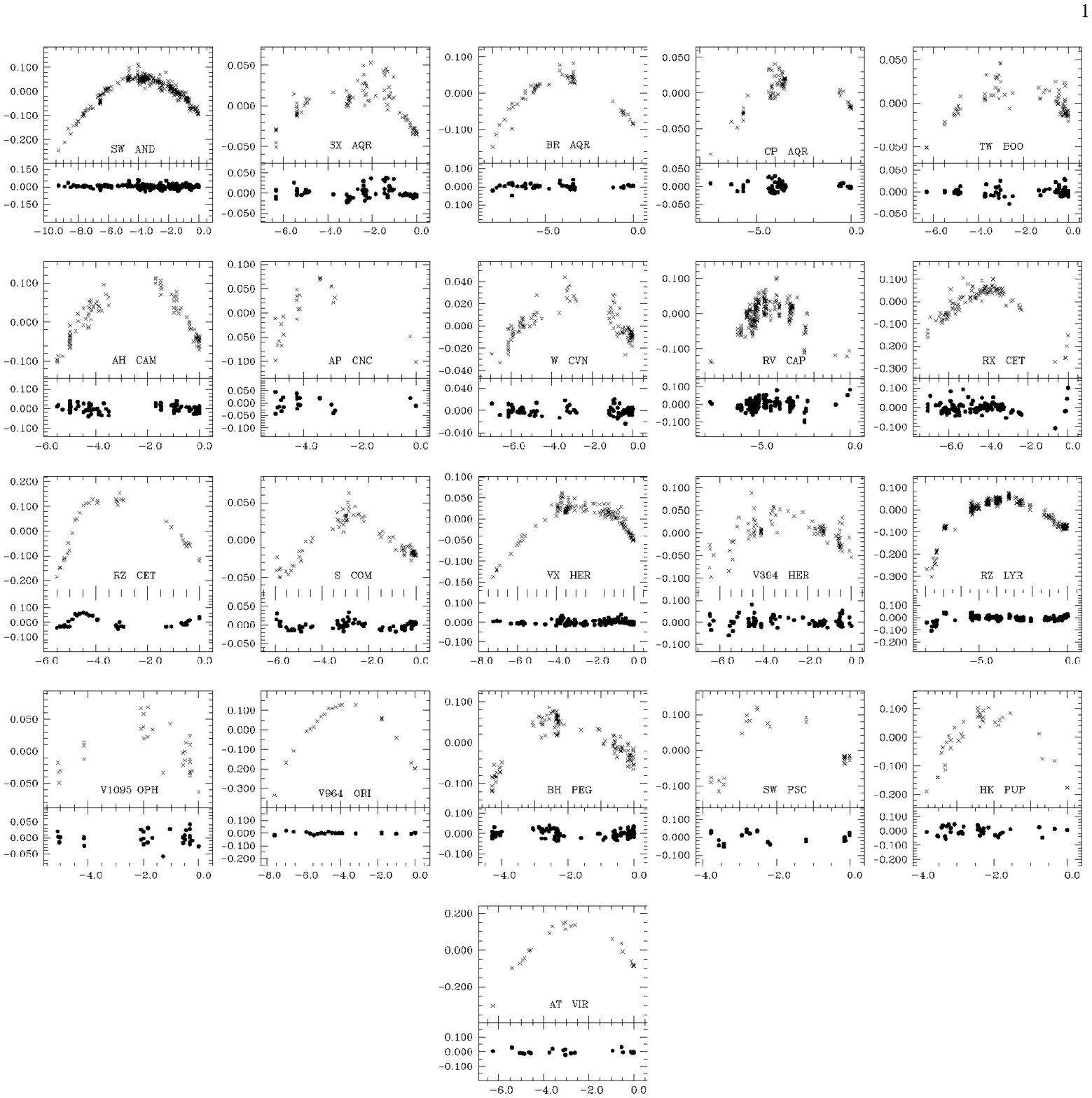}
\caption{\footnotesize O--C values for RRab stars. Filled circles indicate O--C
values calculated from the parabolic fit (period decreasing at a constant rate),
crosses from the linear fit (constant period).
On the $x$--axis we plot the
elapsed cycles (E/10000); on the ordinates the O--C values are in days.}
\label{decf}
\end{center}
\end{figure*}
\section{Data analysis}\label{sect_datan}
We extracted from the GEOS database the RRab stars for which the times of maximum
brightness span more than 50~years. Actually, the observations of RR Lyr stars
started at the end of the XIX$^{\rm th}$ century and for several objects the
time baseline exceeds 100~years, with data in three different centuries.
Throughout this time the observing techniques were very different (visual,
photographic, photoelectric, CCD) and for each technique the accuracy of the
data was not homogeneous, depending on the observer, the equipment and the number
of measurements. Unfortunately, all that information was seldom known in the case
of very old observations (often disseminated in several papers), which in turn
are very important for establishing the O--C trends. In these conditions, the
attribution of weights to the  maxima observed with different techniques was not
practicable in a rigorous way. Moreover,  when viable, this weighted procedure
yielded results not significantly different from those obtained without weights;
indeed, in these cases the datasets were mostly composed of data from a small number
of observers (often using the same technique) and the resulting differences among
weights were small. In  a later stage, we also tried to assign weights to
different subsets considering their scatter from a fitting O--C curve (see
Sect.~\ref{sect_cstrate}), and this procedure also yielded similar results.
Therefore, we are confident that our conclusions are not affected by the
constraint of having assigned the same weight to all the maxima.

Particular care has been taken in the correct evaluation of the cycle values to
be assigned to old maxima or after large gaps. Different O--C plots were obtained
from different cycle countings, and abrupt jumps have been considered as suspicious.
Sometimes, in the presence of large O--C variations we could not reach a uniquevocal
evaluation of the elapsed cycles, and therefore we did not consider the star
for the subsequent analysis. During this process we also corrected some times of
maximum brightness for obvious typographical errors; outliers (perhaps originated
by errors in the decimal numbers) were also rejected, especially within otherwise
well--defined small groups of maxima.

In several cases we corrected the linear elements since the respective O--C values
were linearly increasing or decreasing. The new periods, calculated on a very long
time baseline, provided more reliable O--C values. After this refinement, we started
the analysis of the long--term behaviors. This procedure seems more reliable than
comparing mean periods derived from groups of observations separated by very large
gaps over a very long time baseline, as  happens for globular cluster stars. Of
course, in the case of galactic RR Lyr stars we met large gaps in the data,
but we disregarded unclear cases.

Table~\ref{inven} summarises the inventory of the different patterns we observed
in the O--C plots. Amongst the 123 cases, 81 RRab stars are brighter than 12.5 at
minimum. To evaluate the completeness of our sample, we note that the GCVS lists
164 RRab stars with a magnitude at minimum brighter than 12.5. Thus, our sample
is complete at the 49\% level; of course, there is no way to recover the remaining
51\%, since for these stars an insufficient number of times of maximum is available.
Only in the future, with the progressive monitoring ensured by telescopes as TAROTs,
will it be possible to increase the number of well--observed RRab stars. We note
that amongst the 84 unstudied variables, 56 have $\delta<0^\circ$.

The RRab stars showing a constant period  for which we can provide improved
elements are listed in Table~\ref{cst}. Formal errors at the 1~$\sigma$--level
from the least--squares fits are reported. The RRab stars whose elements cannot
fit all the observed times of maxima are the most interesting cases.
They are described  in the next sections. We note that in spite of
the short periods, \object{RS Boo}, \object{AH Cam} and \object{EZ Cep} are RRab stars.
In particular, the GCVS classification of \object{EZ Cep} (RRc) is wrong; the
{\sc hipparcos} photometry clearly shows a very steady ascending
branch, typical of RRab stars (see Fig.~10 in Poretti \cite{ennio}).
\section {Stars with  a linearly  variable period}\label{sect_cstrate}
The most striking result is the detection of a large number of O--C plots showing
a parabolic pattern when linear elements are used. The cause of this pattern is
a regular period variation. The coefficients of the parabolic fits and their
error bars have been calculated by using standard routines (Press et al.
\cite{press}). Sometimes they are not well defined; as a rule of acceptance the
correlation coefficient $r$ of the least--squares parabola has to be larger than
0.70. The times of maximum have been fitted with the elements
\begin{equation}
C = T_0 + P_0\cdot E + a_3\cdot E^2
\end{equation}
where $P_0$ is the period (in days) at  epoch $T_0$ and $E$ is the cycle counter.
Then, the linear rate $dP/dt$ (in [d/d]) is derived from
\begin{equation}
dP/dt = 2\cdot a_3/ \langle P \rangle
\end{equation}
where $\langle P \rangle$ is the average period, i.e., the ratio between
the elapsed time (the difference between the first and last calculated maxima)
and the elapsed cycles.
In the literature it is more common to indicate the rate of period changes using
the $\beta$ parameter, expressed in  [d~Myr$^{-1}$] (e.g., Lee \cite{lee}, Jurcsik
et al. \cite{omega}).
The $\beta$ values can be calculated from the second--order fits by applying
\begin{equation}
\beta = 0.0732 \cdot  10^{10} \cdot a_3/\langle P \rangle
\end{equation}
It indicates how much the period will change in 10$^6$~years. In turn, this change
can also be expressed as a fraction of the pulsation period
(e.g., Smith \& Sandage \cite{mfifteen}), i.e.:
\begin{equation}
\alpha=\beta/\langle P \rangle
\end{equation}
Table~\ref{tinc} and Table~\ref{tdec} list the parabolic elements for RRab stars
showing an increasing or decreasing period, respectively. With the aim of providing
elements that will be directly useful in the planning of new observations, in
each dataset the E=0 value has been assigned to the most recent time of maximum.
Blazhko variables are also identified. The effect of having introduced the
quadratic term can be evaluated for each star by comparing the lower and upper
panels in Fig.~\ref{incf} and Fig.~\ref{decf}.

Amongst the 27 RRab stars showing a linearly increasing period (Table~\ref{tinc}
and Fig.~\ref{incf}), some cases are at the limit of acceptance ($0.70\le~r~\le0.76$).
In particular, the reliability of the older maxima is decisive in inferring the
parabolic trends for \object{V341 Aql}, while that of the more recent ones is decisive
for \object{EZ Cep}. In all the other cases the parabolic trend is clearly defined; in
particular, $r>$0.85 in  20 cases. The standard deviations of the parabolic fits
are less than 0.025~d in all cases except three (\object{TV CrB}, \object{SV Eri} and \object{TV Leo}).
In particular, \object{SV Eri} is the RRab star showing the largest O--C variation and
the largest $dP/dt$ value, one order of magnitude larger than the average.

Amongst the 21 RRab stars showing a linearly decreasing period (Table~\ref{tdec} and
Fig.~\ref{decf}) the reliability of the older maxima is decisive in inferring the
parabolic trends for \object{TW Boo} and \object{V394 Her}. In the \object{RV Cap} case the parabolic trend
is supported by both old and recent maxima, while the Blazhko effect (see next
section) strongly contributes to reducing the $r$ value (0.69). In 17 cases we
have $r>$0.85.

The standard deviations listed in Table~\ref{tinc} and Table~\ref{tdec} have a
mean value of 0.016~d and a median value of 0.014~d. They can be considered
as an estimate of the accuracy of the photographic and visual times of maximum
brightness, since they constitute the largest fraction of the data.
We can infer that the weighting procedure would not have been very sensitive to
the more accurate (by a factor from 5 to 10) CCD and photoelectric data;
these types of data are still rare and grouped in the last few decades.
This could explain why we obtained similar results by applying the weighting
procedure when possible (see Sect.~\ref{sect_datan}). The above
statistics and comparisons are quite satisfactory, and though the classification
of some cases is somewhat doubtful, in general the observational scenario of stars
with a linearly variable period is very well defined.

Just a few of the stars listed in Table~\ref{tinc} and
Table~\ref{tdec} were known to show a second--order term (\object{BN Aqr},
\object{BR Aqr}, \object{CP Aqr}, \object{RW Cnc}, \object{W CVn}, \object{VZ Her},
\object{RR Leo}, \object{AV Peg}, and \object{AR Per}), and in all cases there is an
excellent agreement between the previous values (see for instance Ol\'ah \&
Szeidl \cite{olah} and Szeidl et al. \cite{szeidl} or, more in general, the GCVS,
Kholopov \cite{Kholopov}) and ours.
\subsection{Other types of O--C  variations}
\begin{figure*}
\begin{center}
\includegraphics[width=1.95\columnwidth]{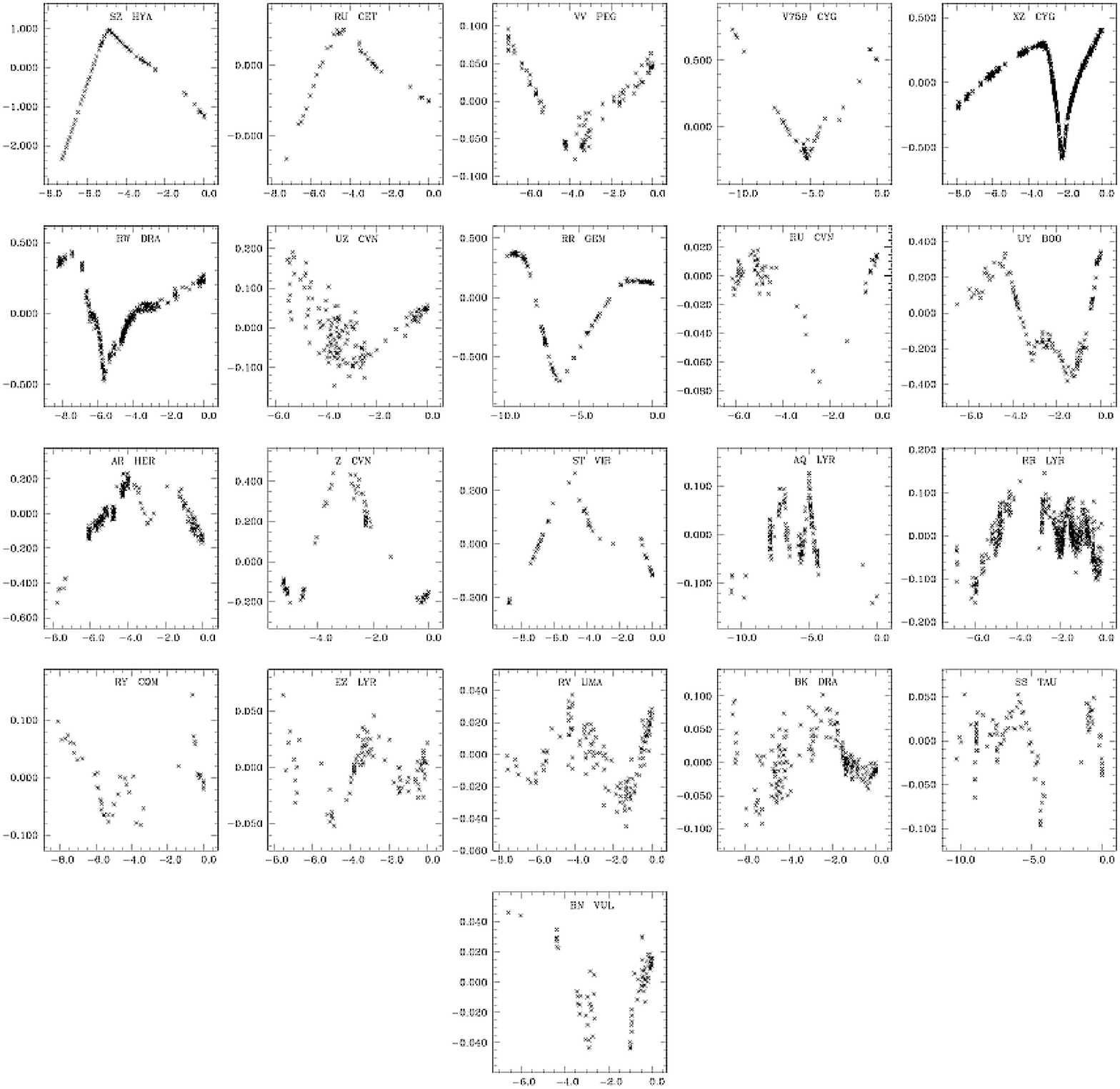}
\caption{\footnotesize O--C values for RRab stars calculated with a linear fit
and showing irregular variations. They are ordered by increasing complexity of
their O--C patterns.  On the $x$--axis we plot the elapsed cycles (E/10000); on
the ordinate one we plot the  O--C values in days.
}
\label{cont}
\end{center}
\end{figure*}

In our sample we can find examples of O--C patterns that are too complicated to
be described with a polynomial fit. The O--C values derived from linear elements
(not listed because they are valid only as rough average values on the considered
time baseline) for several stars show irregular oscillations, with different
behaviors (Fig.~\ref{cont}). We note that in several cases there is no gap in the
observations and therefore there is no ambiguity about the general O--C behavior.
The non--repeating O--C oscillation shapes rule out light--time effects. A Blazhko
effect can be partially responsible for some scatter, but not for the general trend.

The cases of \object{SZ Hya}, \object{VV Peg}, \object{V759 Cyg} and \object{RU Cet}
are the simplest ones, since it
seems that a single, abrupt change in the period has occurred. This also seems
to be the case of \object{UZ CVn}; we rejected the parabolic fit since the left branch
seems steeper than the right one, despite of the large scatter. \object{XZ Cyg}, \object{RW Dra},
\object{RR Gem} and \object{RU CVn} are a little more complicated; the changes occurred more than once,
but still they are very sharp. The complexity of the O--C patterns increases in
the cases of \object{UY Boo}, \object{AR Her}, \object{Z CVn}, \object{ST Vir},
\object{AQ Lyr}, \object{RR Lyr} itself, and \object{RY Com},
with several changes in the direction of the period variations. Other cases
(\object{EZ Lyr}, \object{RV UMa}, \object{BK Dra}, \object{SS Tau}, \object{BN Vul})
show a smaller amplitude of the O--C
values, resulting in more scattered plots. However, the general behaviors are
very similar to the previous ones, with two or more small changes in the periods.
\section{Searching for Blazhko and light--time effects}\label{sect_BLZH}
The most obvious way to verify the reliability of the new elements,
both linear and quadratic, is to analyze the residual O--C values. To do that, we
applied the iterative sine--wave least--squares method (Vanicek \cite{vani}). The
frequency analysis was performed in the interval 0.00--0.20~cd$^{-1}$.

In  several cases (\object{SX Aqr}, \object{X Ari}, \object{RS Boo}, \object{RW Cnc},
\object{S Com}, \object{SU Dra}, \object{RZ Lyr},  \object{AV Peg}, \object{BH Peg}
and \object{RY Psc}) a very long period has been detected. Such a periodicity is
often about half the time coverage and the amplitude is quite small. Therefore,
we can infer that the parabolic fit is not perfect and that some systematic
residuals were left, producing a spurious peak in the subsequent power spectrum.
This feature can be noted looking at the O--C plots of the stars mentioned above
in Figs.~\ref{incf} and \ref{decf}: the two branches seem to be  asymmetric.
We improved the O--C fitting by calculating a third--order least--squares fit,
but we obtained a significant third--order term in three cases only (RS Boo,
RW Cnc, SU Dra). They produce similar rates of the period changes:
1.70$\cdot$10$^{-10}$ d/d (second--order fit) against 1.34 $\cdot$10$^{-10}$ d/d (third--order fit) for SU Dra,
7.08$\cdot$10$^{-10}$ d/d (second--order fit) against 6.19 $\cdot$10$^{-10}$ d/d (third--order fit) for RW Cnc,
1.42$\cdot$10$^{-10}$ d/d (second--order fit) against 1.70 $\cdot$10$^{-10}$ d/d (third--order fit) for RS Boo.
We note that new TAROT observations in the next decade will allow us to disentangle
the O--C behavior in these intriguing cases.

RZ Cet seems to be a particular case. The residual O--C plot is quite unusual;
a long term oscillation is clearly visible and seems to be superimposed on the
parabolic trend (Fig.~\ref{decf}). The frequency analysis detects a low--frequency
term at $f$=0.00008~cd$^{-1}$, corresponding to $P$=12500~d, i.e., half the time
coverage. The possibility of an erratic variation of the period is still plausible,
but here the full amplitude is very large (0.082~d, i.e., 2.0~h). The standard
deviation after the introduction of the long period decreases from 0.034~d to
0.017~d and the fitting curve is perfectly sine shaped. These clues support the
possibility of a light--time effect, but the gap in the O--C coverage hampers
complete certainty.

The presence of a Blazhko effect has been reported for \object{SW Boo}, \object{SW Psc}
and \object{FK Vul}, but available times of maximum light are too scant to perform a
reliable frequency analysis. The frequency analysis of the residual O--Cs related
to \object{SW And}, \object{DM Cyg}, \object{RZ Lyr} and \object{BH Peg} did not
detect any evident peak in the power
spectra, though these RR Lyr variables are reported to be Blazhko stars. However,
in some stars the amplitude of the Blazhko period changes over time, while in
others the Blazhko effect might be more evident in the periodic modulation of
the amplitude than in the O-C values.  Therefore, in both cases the amplitudes
of the O--C shifts could be too small to rise above the noise level. In other
cases the detection has been positive. They constitute a convincing demonstration
of the quality of the database.

\begin{figure}
\begin{center}
\includegraphics[width=0.95\columnwidth]{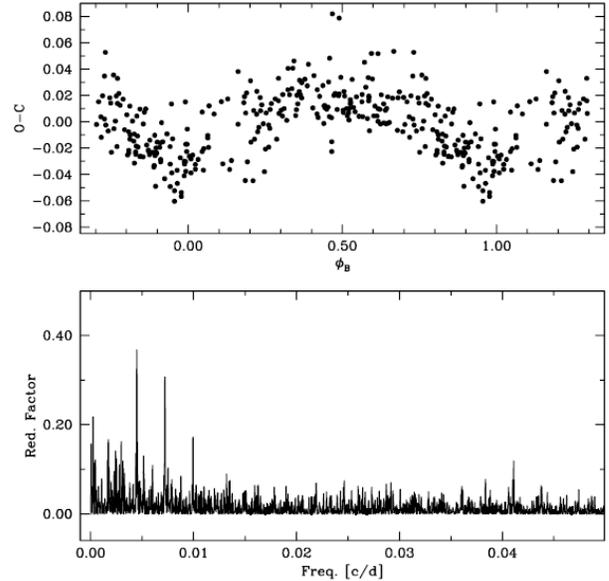}
\caption{\footnotesize The long--period Blazhko effect of \object{RV Cap}. Lower panel:
power spectrum of the O--C values with respect to the parabolic elements;
the highest peak is at 0.00448~cd$^{-1}$. Upper panel: the same O--C values
folded with the Blazhko period of 222.96~d. O--C measure unit is days.}
\label{rvcapbla}
\end{center}
\end{figure}

{\it \object{RV Cap}} -- Kholopov (\cite{Kholopov}) reports a Blazhko period of 225.5~d.
Our frequency analysis refined this value to 222.96$\pm$0.14~d. Figure~\ref{rvcapbla}
shows the power spectrum (lower panel) with the peak at $f$=0.00448~cd$^{-1}$
clearly standing out. The full amplitude of the O--C variation is 0.045~d.
In spite of the heterogeneous observational techniques, the O--C curve appears
quite evident when folding the O--C over the 222.96~d period (upper panel;
T$_0$=2430027.87). The Blazhko effect is responsible for the large scatter observed
in the O--C plot of RV Cap in Fig.~\ref{decf} and for the low $r$ value (0.69).

\begin{figure}
\begin{center}
\includegraphics[width=0.95\columnwidth]{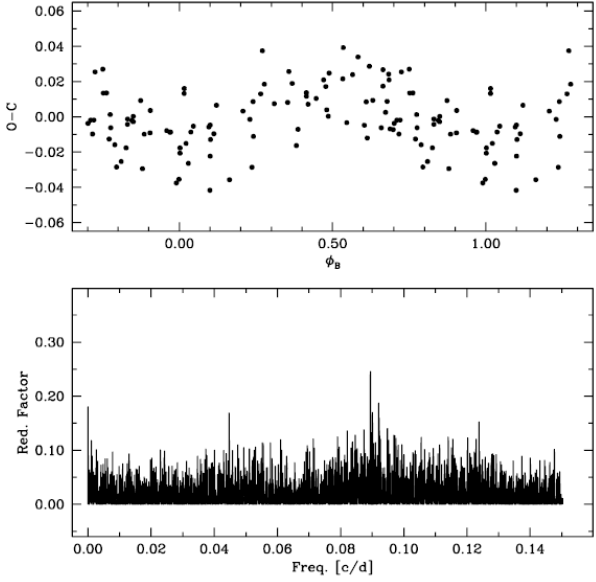}
\caption{\footnotesize The short--period Blazhko effect of \object{AH Cam}. Lower panel:
power spectrum of the O--C values with respect to the parabolic elements;
the highest peak is at 0.0894~cd$^{-1}$. Upper panel: the same O--C values folded
with the Blazhko period of 11.1808~d. O--C measure unit is days.}
\label{ahcambla}
\end{center}
\end{figure}

{\it \object{AH Cam}} -- The Blazhko period of AH Cam is one of the shortest known so far.
Smith et al. (\cite{smith}) identified a period of 11~d by means of two photometric
runs in 1989--90 and 1991--92 (covering 17 and 13 nights, respectively). The light
curves show evident changes in shape and amplitude. The Blazhko effect is also
recoverable in the times of maximum light spanning 56~years (Fig.~\ref{ahcambla}).
The frequency analysis unambiguously identified the peak at 0.0894~cd$^{-1}$ (lower
panel), and the O--C plot over the refined period at 11.1808$\pm$0.0004~d
(T$_0$=HJD~2438040.96) shows a full amplitude of 0.024~d (upper panel).

{\it \object{RS Boo}} -- The power spectrum reveals the low--frequency peak accounting
for the slightly  asymmetric O--C behavior (see above). However, a second peak
is visible at 534.24~d, very close to the value given by Oosterhoff (\cite{oost}),
i.e., 533~d. The full amplitude is 0.010~d. On the other hand, there is no evidence
of the additional period in the 58--62~d range suggested by Kanyo (\cite{kanyo});
its amplitude, if real, is probably too small to be detected.

{\it \object{TT Cnc}} -- The power spectrum shows a peak at $f$=0.01145~cd$^{-1}$,
corresponding to 87.35~d, close to the value determined by Szeidl (\cite{bela}),
i.e., 89~d. The resulting O--C  curve shows some scatter.

{\it \object{CX Lyr}} -- The standard deviation of the O--C values with respect to the
parabolic elements is quite high (0.036~d). The spectral window of the data is
poor and hence the frequency analysis detects several peaks with comparable heights.
The most convincing peaks correspond to periods of 227~d and 128~d.
The O--C values folded over the two periods yield very
similar plots, with a small preference for the 128~d curve; the full amplitude
is 0.070~d and the standard deviation decreases to 0.027~d. No Blazhko effect
was  previously reported on CX Lyr.
\section{Period variations as evolutionary effects}\label{sect_discuss}
\begin{figure}
\begin{center}
\includegraphics[width=0.95\columnwidth]{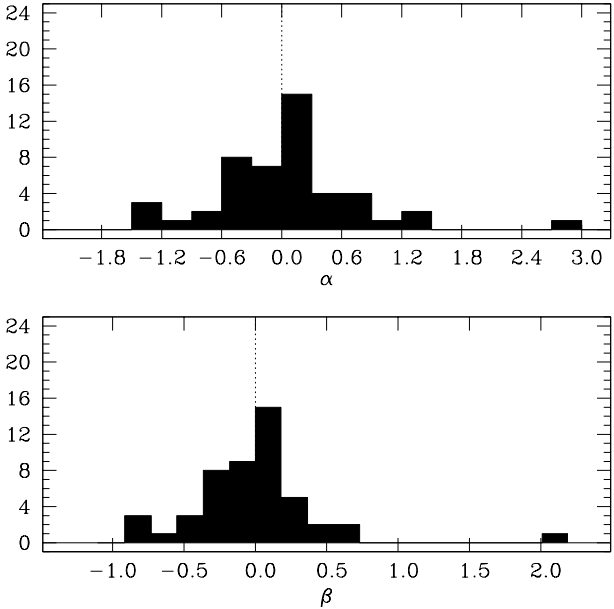}
\caption{\footnotesize Distribution of the rates of period changes. Constant
periods (i.e., $\beta$=0.0) are not shown. Note the extreme case of \object{SV Eri}
in the positive part.}
\label{istorate}
\end{center}
\end{figure}
\begin{figure}
\begin{center}
\includegraphics[width=0.95\columnwidth]{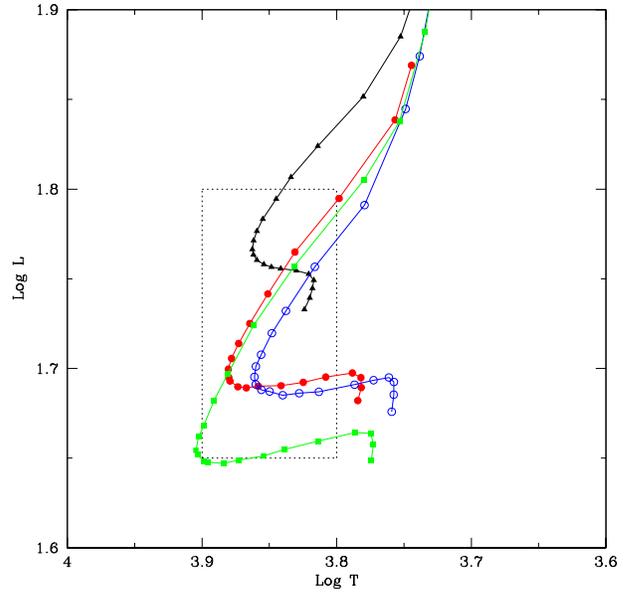}
\caption{\footnotesize Examples of tracks showing blueward evolution at the
beginning of the horizontal branch evolution. Triangles (in black) indicate a
track with $Z$=0.0001 and M=0.86~M$_{\sun}$, filled circles (in red) a track
with $Z$=0.0004 and M=0.74~M$_{\sun}$, empty circles (in blue)  a track with
$Z$=0.0007 and M=0.72~M$_{\sun}$, squares (in green)  a track with $Z$=0.0010
and M=0.68~M$_{\sun}$. The box indicates the instability strip domain (Lee
\cite{vict}).}
\label{tracks}
\end{center}
\end{figure}
\begin{figure}
\begin{center}
\includegraphics[width=0.95\columnwidth]{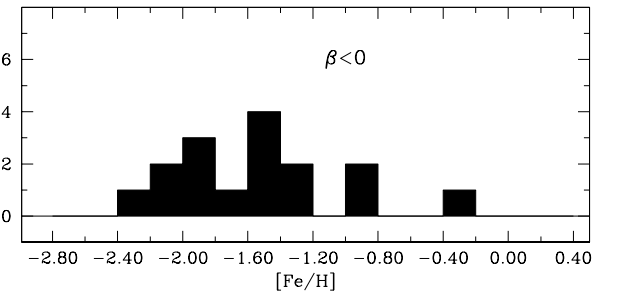}
\caption{\footnotesize The distribution of the [Fe/H] values (as reported by
Layden \cite{layden}) of the stars showing decreasing periods (16 out of 21 cases).
}
\label{feh}
\end{center}
\end{figure}

The $dP/dt$ values listed in Table~\ref{tinc} and Table~\ref{tdec} constitute the
first measurements of the rate of period changes in the sample of the best observed
RR Lyr stars belonging to the Milky Way field. It is noteworthy that all the
values, both positive and negative, have the same order of magnitude and therefore
they should have a common origin. Such an origin cannot be ascribed to a particular
cause (as it could  for  period variations induced by tidal effects or by particular
instabilities), but should be provoked by a quite general phenomenon, involving
a whole population of stars. Though we cannot rule out the possibility that an
individual star
amongst those listed in Table~\ref{tinc} and Table~\ref{tdec} is in a particular
transition phase (see Sect.~\ref{ercha}), the immediate answer is that the period
variations are caused by long--term stellar evolution; indeed, RR Lyr stars
change their radii and in turn their fundamental radial periods when moving along
the horizontal branch. The order of magnitude of the period changes we detected is
that expected from evolutionary  models: Lee (\cite{lee}) reports $\beta$ values
in the range 0.0--0.3~d~Myr$^{-1}$. Moreover, our values are similar to those
reported by Jurcsik et al. (\cite{omega}) in the case of $\omega$ Cen RR Lyr stars
($\beta$ mean rate of 0.15~d~Myr$^{-1}$) and discussed as evolutionary changes.
Therefore, our analysis, providing a reliable estimate of the rate of period changes
in field RR Lyr stars, allowed us to perfom a closer investigation of the stellar
evolution along the horizontal branch.
\subsection{Comparison between stars with period changes of opposite sign}
The histograms shown in Fig.~\ref{istorate} summarise the properties of our sample.
They provide evidence that a large fraction of  RR Lyr stars experience
appreciable changes in their structure in the time now covered by observations.
The first immediate result is that the blueward evolution (decreasing periods)
is as common as the redward one (increasing periods). Consequently, RR Lyr stars
seem to be an excellent laboratory to verify the modifications of the internal
structure caused by stellar evolution under different conditions (e.g., contraction
and expansion). On the other hand, we remind the reader that in 54 cases (i.e.,
a number of cases similar to the number of RR Lyr stars showing period variations)
we did not detect a significant period variation. No period variation could occur
at the turning point in the evolutionary tracks, but a constant period probably
means that  there are phases in the instability strip crossing when
changes are not large enough to be measurable in 100~years or so.

When limiting to cases where a period variability is noted, we get a similar
distribution between negative and positive rates (Fig.~\ref{istorate}) for our
field  RRab stars. We also note that the period ranges are nearly coincident in
the two samples; in particular, we have a mean and median value of 0.53~d for
stars with increasing period, and a mean and median value of 0.51~d for stars
with decreasing period. This similarity is not observed for RRab stars in galactic
clusters (Lee  \cite{lee}, Jurcsik et al. \cite{omega}), where positive rates
are largely more frequent.

In our opinion, the large number of galactic stars showing decreasing periods
is a strong confirmation of the evolutionary tracks provided by Sweigart
(\cite{tracks}) and Lee \& Demarque (\cite{ldemarque}). Most of these tracks
cross the instability strip at high luminosities in the redward direction. However,
Fig.~\ref{tracks} shows that for some combinations of masses and metallicities
(e.g.,  by increasing the mass of the model with a lower metallicity), a blueward
evolution at the beginning of the horizontal branch phase is  predicted, both at
high luminosity ($Z$=0.0001 and M=0.86~M$_{\sun}$) and at low luminosity ($Z$=0.0010
and M=0.68~M$_{\sun}$), and such blueward paths actually cross the instability
strip domain. The theoretical sequences calculated by Dorman (\cite{dorman})
predict a blueward evolution at slightly lower luminosities, crossing the instability
strip in fewer cases (e.g., Fig.~7 in Jurcsik et al. \cite{omega}). In general,
all theoretical models predict a blueward evolution in connection with the prevalence
of the H--shell burning over the He--central burning. The fact that the intersection
between these blueward tracks and the instability strip is predicted only for
particular combinations of physical parameters is compensated by the longer lifetimes
on the blueward path relative to the redward one (see next section). Therefore, once
we have established that the blueward crossing occurs, the slow migration
could explain why we observe RRab stars in this phase.

To investigate the real agreement between the input parameters of the evolutionary
tracks shown in Fig.~\ref{tracks} and the stars of our sample, we considered the
[Fe/H] values determined by Layden (\cite{layden}) by means of homogeneous
low--to--moderate dispersion spectra. Sixteen stars with decreasing periods are
included in Layden's sample; we get [Fe/H]$<-1.20$ in 13 cases and a median value
of [Fe/H]$=-1.48$ (Fig.~\ref{feh}). These [Fe/H] values translate into $Z\leq0.0010$
and $Z=0.0007$ (e.g., Pietrinferni et al. \cite{basti}). Therefore, the observed
metallicities are in the same range of those used to generate the tracks shown
in Fig.~\ref{tracks}. A similar distribution is obtained for stars with increasing
periods; in 20 cases out of 24 we observe [Fe/H]$<-1.20$ and the median value is
[Fe/H]$=-1.52$. This supports the hypothesis that all the  stars with
increasing and decreasing periods have similar characteristics, but the latter
ones are less evolved than the former.
\subsection{The rate of period changes}
To avoid influence from a few very large positive or negative rates, we calculated
the median values of the rates of the period changes to characterise the two
distributions shown in Fig.~\ref{istorate}. We obtained $\beta=+0.14$, $\alpha=+0.26$
for the 27 stars with increasing periods and $\beta=-0.20$, $\alpha=-0.43$ for
the 21 stars with  decreasing periods. As already shown in early investigations
by Sandage (\cite{pion}) and then confirmed by Lee (\cite{lee}) and Pietrinferni
et al. (\cite{basti}), the estimate of the horizontal branch lifetime is about
$\sim$10$^8$~years, and a star spends $\sim$1/3 of its life in the instability strip.
Moving uniformly through the strip with time, the corresponding rate of period
change should be around $\alpha$=0.02: Fig.~\ref{istorate} provides evidence of
rates one order of magnitude greater.

However, we have to take into account that when observing a star for 100 years
we are monitoring a fraction corresponding to $10^{-5}$ of the lifetime spent
crossing the strip. The rates of the period changes we measure are a sort of
instantaneous value, which refers to a particular stage in the horizontal branch
phase. Indeed, it is quite obvious that the stars cannot maintain  a constant
rate of some 0.1~d~Myr$^{-1}$ for several 10$^6$~years, since their periods would
assume values largely outside the RRab range (0.4--0.8~d) and thus physically
meaningless. Therefore, we should definitely refrain from considering the observed
rates as the constant rate of the instability strip crossing of a specific stellar
model. Such large variations (units of 0.1 d~Myr$^{-1}$) must last for a few
10$^6$~years only. On the other hand, the 54 stars showing a constant period could
be in the middle of their long crossing of the instability strip.

The obvious interpretation of Fig.\ref{istorate} is to consider that the observed
rates are those corresponding to the phases when the stars change their periods
with the fastest rate. The rate spread is caused both by the different physical
parameters of the stars and by the small differences in age. When considering
the evolutionary rates provided by Lee (\cite{lee}), it seems that large $\beta$
rates occur only at the end of the instability strip crossing (i.e., roughly,
1/20 of its horizontal branch lifetimes). However, such a proportion would imply
a huge number of RR Lyr stars with nearly constant periods and a few stars with
large $\beta$ values. In our sample we found a 2:1 ratio. To explain the low number
of stars with constant period we should admit that amongst the galactic RR Lyr stars
there is a predominance of low--mass stars: the theoretical models predict (see
Fig.~1 in Lee \cite{lee}) that these stars cross the instability strip very quickly,
close to the end of their horizontal branch lifetime. The case of \object{SV Eri} can
support this point. It has a long period (0.71~d), close to the observational limit
of RR Lyr periods. The theory predicts a large positive rate in this case and
indeed we observe $\alpha$=2.97, actually a huge value (see below). XX And has
a similar period value (0.72~d), but its rate of period change is one order of
magnitude smaller ($\alpha$=0.20). Thus, SV Eri seems to have entered the rapid
final phase, and it should be considered a very peculiar object experiencing a
rapid final evolution.

The median values  also suggest that galactic RR Lyr stars experienced slightly
faster period decreases than increases.  To evaluate the physical meaning of this
fact, we need to compare them more closely with the lifetimes of the blueward and
redward crossings. Actually, the blueward crossings are about 1.6~times longer
than the redward.
As an example, the two tracks with $Z$=0.0004 and $Z$=0.0007 in Fig.~\ref{tracks}
(i.e., those indicated by filled and empty circles, respectively) yield 33 and
28~Myr  for the blueward evolution and  21 and 17~Myr for the redward evolution.
Moreover, the stars should change their radius to a lesser extent in the blueward
evolution than in the redward one (0.13 against 0.20, in $\log~R$). Therefore,
at first glance, the positive $\beta$ values should be on the average greater
(also in absolute value) than the negative ones, since a larger period variation
occurs in a shorter time. This expectation seems to be contradicted by the
histograms shown in Fig.~\ref{istorate}.
The contradiction could be partially explained by taking into account the
observation that the short lifetimes  on the final part of the redward evolution
strongly reduce the possibility of observing these stars. Nevertheless, the
RRab stars showing rapid period decreases still need more detailed
theoretical investigations. For example, their large number could result from a burst
in the stellar formation process in our galaxy.

Finally, we have to address a possible discrepancy between the  observed and
predicted rates of the period changes.  The theoretical evolutionary rates
calculated by Lee (\cite{lee}) and recalled above predict $\beta$ values smaller
than 0.30 for almost all the stars. $\beta$ values larger than 0.30 (stars with
very low mass, $M=0.68~M_{\sun}$, and well--defined metallicity, $Z$=0.0004) are
only reported in exceptional cases and the lifetimes of these phases are very
short. In our sample we have 18 stars with $0<\beta<0.30$ and 9 stars with
$\beta>0.30$. Also in this case, the  2:1 ratio seems too small to be caused by
the presence of exceptional cases. Furthermore, the $\alpha$=2.97 value of SV Eri
recalled above is much higher than expected. Smith \& Sandage (\cite{mfifteen},
also quoting Iben \cite{iben}) indicate $\alpha$=0.5 as an extreme value, but
such a value is not an uncommon one (see Fig.~\ref{istorate}) and is much smaller
than the SV Eri value.

Our conclusion is that the theoretical models should be fine tuned to match
the observed rates in a better way, now that the extensive GEOS database allows
us to measure them in a reliable sample.
\subsection{Erratic changes}\label{ercha}
During the final phase of the helium core burning it is possible that instabilities
occur and that period variations are not regular, both around  a constant value
or superimposed on the general trend (Sweigart \& Demarque \cite{random1}). This
possibility has been suggested by Sweigart \& Renzini (\cite{random2}) and has
been invoked by Lee (\cite{lee}) to explain the few cases of negative values in
the rate of period changes. RR Lyr stars showing blueward evolution have been
considered as suspicious, invoking period changes caused by temporary
instabilities, but we are providing evidence that the number of RR Lyr stars
subject to blueward evolution is similar to the number of stars subject to the redward
evolution. Therefore, the erratic changes  should be considered as representative of
a particular evolutionary stage, as the final helium core burning could be, but
not as an ad hoc solution to explain negative rates only. We also stress that
the erratic changes occur in both directions (see Fig.~\ref{cont}). In our opinion,
the sudden, sharp period changes occurring in the cases similar to SZ Hya could be
the result of particular temporary instabilities, which can also produce sudden
blueward excursions (Sweigart \& Renzini \cite{random2}).  We also
stress that only a comprehensive analysis performed on a large database can avoid
overlooking these sudden changes as evolutionary ones.
\section{Conclusions}
The analysis of O--C variations over a timescale of 100 years or more has proved
to be a powerful tool for providing quantitative tests of the stellar evolution of
the horizontal branch stars. We can stress some well--established observational
facts:
\begin{enumerate}
\item RRab stars experiencing blueward evolution (i.e., period decreases) are quite
common, only slightly less so than RRab stars experiencing redward evolution (i.e.,
period increases). The period ranges covered by the two groups are very similar
and the mean and median period values are nearly coincident;
\item the absolute values of period changes are larger than expected.
Also in the extreme case of the rapid--evolving star \object{SV Eri} the rate is much
larger than expected;
\item the O--C behavior can be very complicated in some cases, with abrupt or
irregular period variations, rather than monotonic. The regular variations caused
by the light--time effect (and hence duplicity), often invoked to explain large O--C
excursions, are not convincingly observed  in our extensive sample.  The physical
explanations should be searched in the stellar structure; \item Blazhko effect
is often superimposed on secular changes, but the monotonic trend due to
evolutionary variations still remains visible.
\end{enumerate}
As a general conclusion about the comparison between our observational results
and the theoretical predictions, we claim that there is a very powerful feedback
between the two approaches. In particular, theoretical investigations should take
into account that we have observational evidence of many RR Lyr stars showing
blueward evolution. The theoretical models should also match the observed $\beta$
values in a more satisfactory way, as these seem to be higher than expected, both
for redward and blueward evolutions. On the other hand, the observational effort
to monitor RR Lyr stars should be continued, possibly extended to stars at fainter
magnitudes, and the accuracy of the maximum time determinations should be improved
(by using automated telescopes), thus obtaining the same information on a shorter
time baseline.
\begin{acknowledgements}
The outlines of this project have been sketched during several GEOS meetings,
where the different knowledge of amateur and professional astronomers found
a very profitable synthesis. The active participation of N.~Beltraminelli, M.~Benucci,
R.~Boninsegna, M.~Dumont, J.~Fabregat, F.~Fumagalli, D.~Husar, A.~Manna, J.C.~Misson,
J.~Remis, and J.~Vialle to these discussions is gratefully acknowledged.
The authors also thank M.~Catelan, S.~Degl'Innocenti, H.~Smith, and L.~Szabados for useful
comments on a first draft of the manuscript.
EP acknowledges support from the INAF project 39/2005.
The present study has used the SIMBAD database operated at the {\it Centre de
Donn\'ees Astronomiques (Strasbourg)} in France.
\end{acknowledgements}

\end{document}